# Renewable Energy Integration in Distribution System – Synchrophasor Sensor based Big Data Analysis, Visualization, and System Operation

A Dissertation

Presented to the Faculty

of School of Engineering and Computer Science

University of Denver

in Partial Fulfillment

of the Requirements for the Comprehensive Exam

of Doctor of Philosophy

by

Yi Gu

June, 2018

Advisor: Jun Jason Zhang

Author: Yi Gu

Title: Renewable Energy Integration in Distribution System – Synchrophasor Sensor based Big Data Analysis, Visualization, and System Operation

Advisor: Jun Jason Zhang

Degree Date: June, 2018

# Abstract

With the increasing attention of the renewable energy implementation in distributed power system, the hybrid smart grid (SG) operation is mainly featured by distributed renewable generation, data visualization, data prediction in high accuracy and operation cost minimization.

Due to the large volume of heterogeneous data provided by both the customer and the grid side, a big data visualization platform is built to discover the hidden useful knowledge for smart grid (SG) operation, control and situation awareness. An open source cluster computing framework based on Apache Spark is considering to discover the hidden knowledge of the bag-data. And the data is transmitted with a Open System Interconnection (OSI) model to the data visualization platform in a high-speed communication architecture. Google Earth and global geographic information system (GIS) are used to design the visualization platform and realize the result with the test bench.

A short-term load forecasting approach is designed based on support vector regression (SVR) to provide a high accuracy load forecasting for the network reconfiguration. The nonconvexity of the three-phase balanced optimal power flow is relaxed to a optimal power flow (OPF) problem with the second-order cone program (SOCP). The alternating direction method of multipliers (ADMM) is used to compute the optimal power flow in distributed manner. The proposed network reconfiguration is solved in parallel manner with the limited switches and the strong computing capability.

Further more, a multi-timescale operation approach is modeled with a three-phase distributed system, which consists the hourly scheduling at substation level and the minutes power flow operation at feeder level. The system cost with renewable generation is minimized at the substation level. The given error distribution model of renewable generation is simulated with a chance constraint, and the derived deterministic form is modeled with Gaussian mixture model (GMM) with genetic algorithm-based expectationmaximization (GAEM). The system cost is further reduced with the OPF in real-time (RT) scheduling. The semidefinite programming (SDP) is used to relax the nonconvexity of the three-phase unbalanced distribution system into a convex problem, which make sure to achieve the global optimal result. In the parallel manner, the ADMM is realizing getting the results in a short time.

# Acknowledgements

First of all, I would like to extend my sincere gratitude to my supervisor, Dr. Jun Jaon Zhang, for he instructive advice and useful suggestions on my thesis. I am deeply grateful of he help in the completion of this thesis. I learned a lot knowledge from his excellent performance in teaching and research capabilities as well as high motivation and responsibility to his students.

I also gatefully thank Dr. Gao, Dr. Yingchen Zhang, Dr. Eduard, whose profound knowledge of power system helped me a lot in my researches.

I am also deeply indebted to all the other tutors and teachers for their direct and indirect help to me.

Special thanks should go to my friends who have put considerable time and effort into their comments on the draft.

Finally, I am indebted to my mum for their continuous support and encouragement.

# Contents

# List of Tables

# List of Figures

Dedicated to my mum

# Chapter 1

# Introduction

In the recent decades, the increasing demand of energy world needs a rapid development of modern power systems in corporation with the renewable energy. There are still many challenges in the future development of the intelligent distributed power system with the uncertainty of renewable energy. Due to the advantages of green energy, such as economic benefits, little or no pollutions, less maintenance than transitional generators and sustainability, a hybrid distributed power system is still a top topic in the recent power energy researches. However, it is difficult to generate the quantities of electricity as large as the transitional generators, and the renewable energy is a highly weather depended resource. Due to the disadvantages, a transitional distributed power system combined with the renewable energy generation can be satisfactory to both taking the advantages and avoid the disadvantages of renewable energy. [1]

The transitional power system is dramatically changing with the rapid technology development, the new inventions such as smart sensors, large capacity batteries, electric vehicle helps to make the life easier. A large amount of the massive heterogeneous data are provided and indispensable for the SG operation. The way on how to choose the useful hidden information without spatial information is proposed as an open source cluster framework, which is based on Apache Spark and can be used to effectively collect, store and process the data in parallel manner. The discovery knowledge is visualized with a Google Earth Platform and realize interaction with the operators.

The network reconfiguration is used to achieve operating the power system, such as keeping balance of the system load with limited switches controlling, maintaining the voltages profiles and reducing the system loss. An improved method on load prediction technology is used to compute the load deviations in the two scheduled time points, which is always ignored by tranditional ways. The dynamic and efficient network reconfiguration method is simulated based on support vector regressive (SVR). The short-term load forecasting approach is employed to minimize the system loss effectively.

The distributed system markets are attracting more attention in the recent research along with the rapid development technology of the power system. Considering the hybrid power system with high penetration of the renewable energy, the market efficiency and the grid reliability of the combined power system become a new challenge. Due to the economic benefits of the renewable energy, a multi-timescale operational approach is scheduled on substation and feeder level. Different from the transitional computation on the distribution system, a three-phase unbalanced distribution system is used to minimize the system loss. It should not be ignored with a desire of high accuracy on the system loss minimization. And the corrective action is implemented at the substation level and realize the inaction between the customers and the utilities. Reselling the redundant power to real-time power system should be helpful to the operation cost optimization.

## 1.1 Current Research on Knowledge Discovery and Data Visualization

A lot of machine learning related approaches are implemented in power system and related areas such as [2–13]. Similarly, optimization is a large area, which contains both convex optimization and non-convex optimization approaches such as [14–21]. Building a smart grid (SG) with renewable energy generation is imperative, such as solar energy, wind power, hydropower, geothermal, bio energy, etc. [22–24]. Along with the rapid development of the modern power generation technology, the new inventions are dramatically improving the traditional power systems [24–27]. The widely located smart sensors always provide heterogeneous massive data, which are significant indispensable to the smart grid operation and management. However, the useful knowledge always hides in the big amount of the collected data with less spatial information. Due to this, an open source cluster computing framework based on the Apache Spark is built to effectively collect, save and process the big data in parallel manner. And the processed data is effectively transmitted to the proposed visualization platform. At present, a lot of advanced technology companies benefit a lot from the big data development company [28]. For example, the way to improve the system efficiency and extend the service life to determine the optimal position of wind turbine [29]. According to the analysis of the market research, the service market grew from $3.2 billion in 2010 to $16.9 billion in 2015 [30].

The big data in smart grid can be featured with low value density, large volume and high velocity [29, 31–33].

1. Low value density: the necessary and right information always hides in the big amount of the data. For example, in smart grid abnormal data detection, most part of the collected data from operation are general data, the abnormal data only occupy very small parts [34].

2. Large volume: the various smart sensors and high speed sampling rate always provide a large volume of data, which can increase the TB level up to PB level [31].

3. High velocity: the sampling rate speed of the smart sensors are increasing. For example, the synchrophasor measuring device, which measures from 20 samples/s to 1440 samples/s and indicates the fast disturbances in smart grid can be recorded [34].

According to the characteristics, the proposed platform of bag data is described to sat-isfy the requirements as follows:

1. Efficient, resilient and distributed data storage skill for the collecting heterogeneous massive data.

2. Error tolerant, high speed, big data processing and analyzing in parallel manner.

3. Streaming processing platform for high speed in real-time data.

4. Hidden knowledge discovery based on machine learning to meet the requirements of different complex projects.

Hadoop is a widely used open-source software, which is released by Google file system in 2003. It is designed for big data processing and distributed storage in computer clusters [35, 36]. Comparing with the traditional approach, mapreduce is the critical feature of Hadoop, it can divide the job into several smaller jobs and deal with them in distributed computer clustering. In each iteration, reloading the data from disk can cause a longer time consumption, especially for the iterative parameters optimization with machine learning [36]. Based on the new cluster computing framework, Apache Spark is design to reduce the time-consumption, and achieve more resilient on the clusters computation. The important advantages of Apache Spark is *resilient distributed dataset* (RDD) are as follows [35]

1. Using the memory instead frequently reload the data from disk for iterative calculations in machine learning.

2. Dividing the task job into several smaller jobs and compute the results in distributed computer clusters.

3. Rebuilding the information for high resilient error tolerant, especially in iterative analysis.

4. Easy to use in many languages and different operation systems.

At present, Google Earth is a widely used and can be operated in multi different operation systems such as Mac OS, Linux, Windows [37]. A Google Earth platform is described to achieve the data visualization with discovered knowledge.

The visualized platform is designed to meet the requirements of independent system operators (ISOs), customers and the power generations, whhich consists: 1) decision support, 2) forecasting assessment 3) operation and control, 4) security and detection,.

## 1.2 Current Research on Load Forecasting on SVR

The network reconfiguration is an easy way to decrease the system loss, control the voltage stability, and trade off the system load by controlling the limited mount of switches [38–41]. The load prediction technology provides the load deviations between two decided time points, which is always ignored in the traditional network reconfiguration methods. Considering the stochastic load deviation in distributed power system, it is important to set up a efficient and dynamic network reconfiguration approach [27, 42–46]. A support vector regression (SVR) is used to cooperate with the network reconfiguration based on short-term load prediction method to minimize the system loss.

The most parameters in SVR can be solved in a convex manner, several parameters, defined as hyper-parameters, cannot be determined in a same way. The optimization of the parameters are really indispensable of SVR, and dramatically influence the performance and efficiency of the forecaster [47–50]. Due to this, a two steps parameters optimization approach is proposed with grid traverse algorithm (GTA). According to the distributed computation frameworks in [51–54], the hyper-parameter optimization of SVR is described with mapreduce to optimize the parameters in parallel manner, furthermore, reduce the calculation time consumption.

Many heuristic methods are used to solve this problem with a three-phase balanced distribution system. However, the results are easily impacted with the initial points and dropped into a local minimums [26, 27, 55]. To avoid this shortage, a distributed net-work reconfiguration approach is proposed for the distribution power systems with convex relaxation. A second-order cone program (SOCP) based approach is used to relax the three-phase balanced power flow problem [56, 57]. After that, the alternating direction methods of multipliers (ADMM) supplies a distributed approach to solving the relaxed convex optimal power flow problem [58]. Along with the dual ascent alternating and decomposition, ADMM solves the relaxed network reconfiguration problem with a short convergence time consumption. A parallel searching manner is used to traverse all the permutations, and decide the best configuration to minimize the system loss.

## 1.3 Current Research on multi-timescale distributed power system with two levels

With the new development of the power generation technology, the attention of the distribution system markets is increasing in recent researches [59–61]. Considering the output of the high penetration renewable energy generation, the distribution power system faces a big challenge. Many approaches are used to improve the net reliability. In [59], a day-ahead market energy auction is built for distribution system operation. In [62, 63], a multi-timescale power system operation approach is proposed for renewable energy generation. In [64], a storage-based operation approach is used to decrease the operation cost of a distribution power system with renewable energies. In [60, 65], the optimal operation methods are proposed with economic demand response in distribution system with renewable energy. However, all these studies ignored the system loss, which would significantly influence the operation cost. A multi-timescale operation approach is proposed to reduce the operation cost. A three-phase unbalanced OPF is built and integrated to further reduce the system loss.

In [66], the stochastic programming optimization (SPO) provides many benefits to transmission systems. The distribution of the forecasting errors is given to generate the error models, which are formulated as a chance-constraint for SPO in distribution systems. It is assumed that there are more than one renewable generators operating on certain feeders. The aggregated error distribution usually includes multiple Gaussian models, which can be modeled as the Gaussian mixture model (GMM) [67]. The variant Gaussian models indicates the uncertainties of the renewable energies, the genetic algorithm-based expectation maximization (GAEM) is applied for GMM to determine the amount of Gaussian models automatically [68]. According to the this, the chance-constraint of the renewable generation is formulated into a deterministic form for further reducing the computation burden [66].

In [69, 70], the variability of the renewable generation is operated by hourly demand response with day-ahead scheduling, which is scheduled without considering the stochastic net load deviation within an hour. In [71], a single line model instead of three-phase system is used to compute the system loss, which ignores the three-phase unbalanced configuration in distribution power systems. A developed approach is introduced in [72], and a secondary-order cone programming (SOCP) is used to relax the OPF for the three-phase balanced distribution system. Because of the nonconvexity of the three-phase unbalanced OPF in distribution system, the heuristic based approach are proposed in [73], however, this method hardly avoids to fall into the local minimums with a high computation time-consumption for a large-scale distribution systems [74]. To avoid the issues above, an inequality constraint is proposed based on semidefinite programming (SDP) to relax the three-phase unbalanced OPF problem. After that, the proposed formulation for system loss is solved in parallel manner with alternating direction methods of multipliers (ADMM). It helps to further reduce the total operation cost hourly.

The **main contribution** of the study is that, for a distribution system operator, a multi-timescale approach is proposed based on a three-phase unbalanced distribution system. With the day-ahead dispatch of the substation level, the objective is provided to optimize the system cost with renewable energy.. The chance constraint is build based on GMM and GAEM. In the feeder level, the objective focus on minimizing the system loss, an OPF problem is formulated for the three-phase unbalanced system, get the global result with ADMM. In the proposed multi-timescales model, the feeder scheduling is described with higher time resolution and update frequencies, which can be obtained near RT calculation.

## 1.4 Architecture of the Paper

In Chapter 1, the literature review of the current research around renewable energy distribution power systems are described, which provides background study of the renewable energy distributed power system for the paper. In Chapter 2, Knowledge discovery in a massive data sets is carefully studied and the useful hidden data is transmitted into a visualized platform, the result provide the audience a convenient way to get the information. In Chapter 3, the network reconfiguration is used to operate the power system with an improved load forecasting technology based on SVR. In Chapter 4, a multi-timescale operational approach is used on a two level distributed power system, which consistes the substation and the feeder level. The numerical results are simulated on IEEE 123-bus, IEEE-8500 and University of Denver campus distributed power system. In Chapter 5, the conclusion and the future work is summarized.

# Chapter 2

# Knowledge Discovery and Data Visualization of Smart Grid Operation

## 2.1 Architecture of the Platform

### 2.1.1 Main Architecture

The main structure of the paper is descried in Fig. 2.1. The proposed platform has four major applications which consists of 1) forecasting and risk assessment, 2) decision support and management, 3) operation and control, 4) security and protection. The power plants, independent system operators (ISOs) could ask the useful knowledge from the proposed platform, respectively.

### 2.1.2 Detailed Architecture

As shown in Fig 2.2. Data collection, resilient distribution dataset, Spark master, GIS visualization, and Applications.

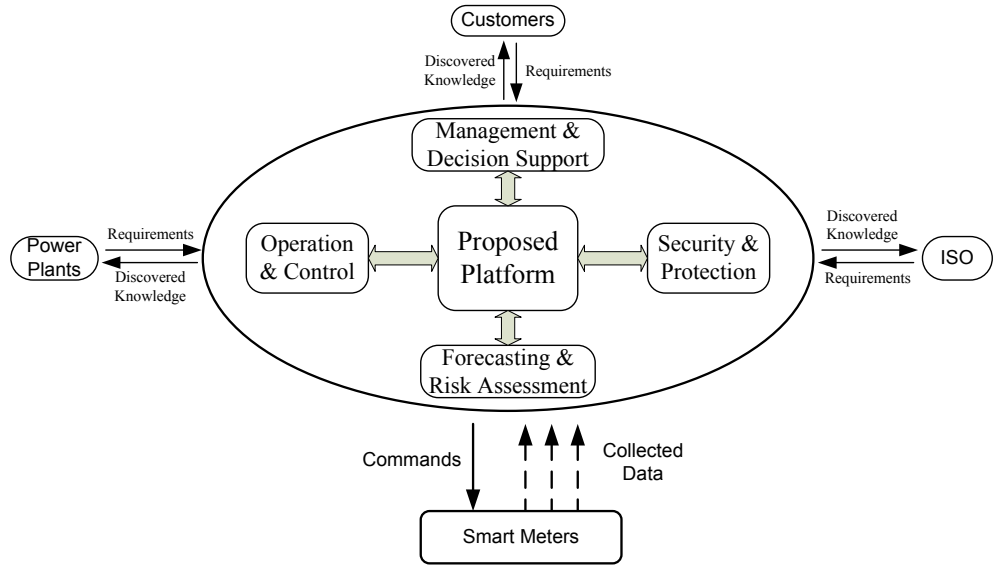

Figure 2.1: Main architecture of the proposed big data platform.

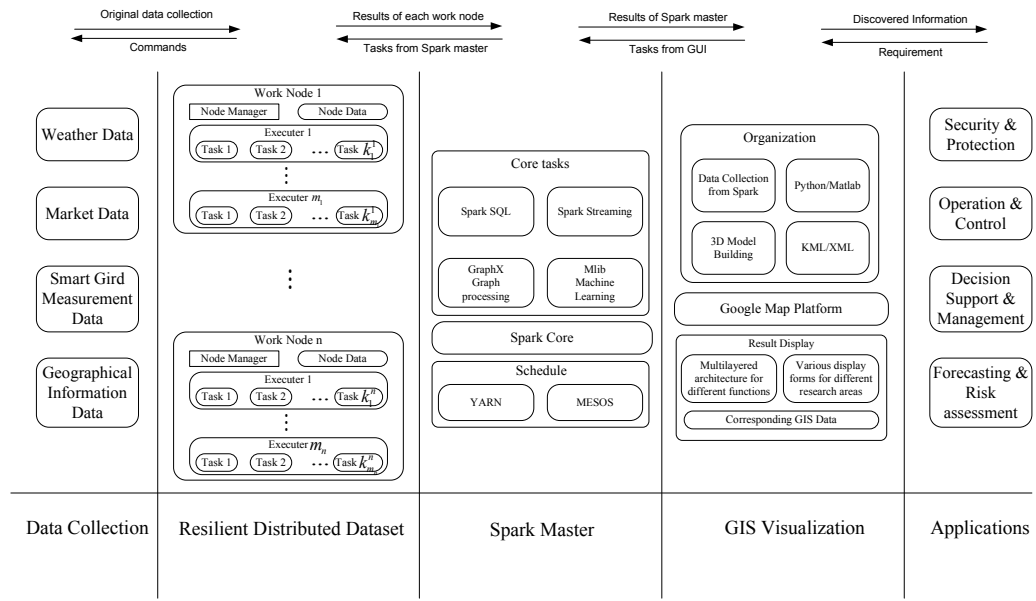

Figure 2.2: Detail architecture of the proposed big data platform.

1. **Resilient Distributed Dataset**: the major task is to store and process the data with the supervising of Spark master. On the one hand, the work nodes from 1 to $n$ are working in parallel. On the other hand, the distributed datasets receive the original dataset and store the dataset in distributed data centers, it can reduce the communication burden to the Spark master. In addition, the critical data can be compressed and stored in distributed to increase the resilient of the database [34].

2. **Data collection**: the major task is to collect different types of data in natural world. In this part, the smart sensors provide the original data from natural world and the smart sensors can receive commands from the Spark master about measurement time periods.

3. **GIS visualization**: the data flow of the discovered knowledge is from the Spark master, the 3D models, Python and Matlab are used to generate the KML files for visualizing data in various forms. Based on different kinds of scenarios, Multi-layered architecture is applied, and different kinds of display forms are used for different research areas.

### 2.1.3 Data Communication Architecture

The architecture of the communication system in the proposed platform is described in Fig 2.3. The two major layers are: data flow in lower layers, application in upper layers. The standard Open System Interconnection (OSI) model is used to compare and explain the corresponding abilities and applications of each layer in the detail architecture.

1. **Network Access**: The major task of this layer is to build a reliable physical wired or wireless communication connection among the work nodes for transmitting and receiving data. In the data collection part, as shown in Fig 2.2, the smart sensors are pervasively located, both wired and wireless communication connections can be implemented according to the data volume and transmission speed. In the part of resilient distributed dataset, Spark master, GIS visualization, because the data volume is large and transmission speed is very high, the wired communication is preferred.

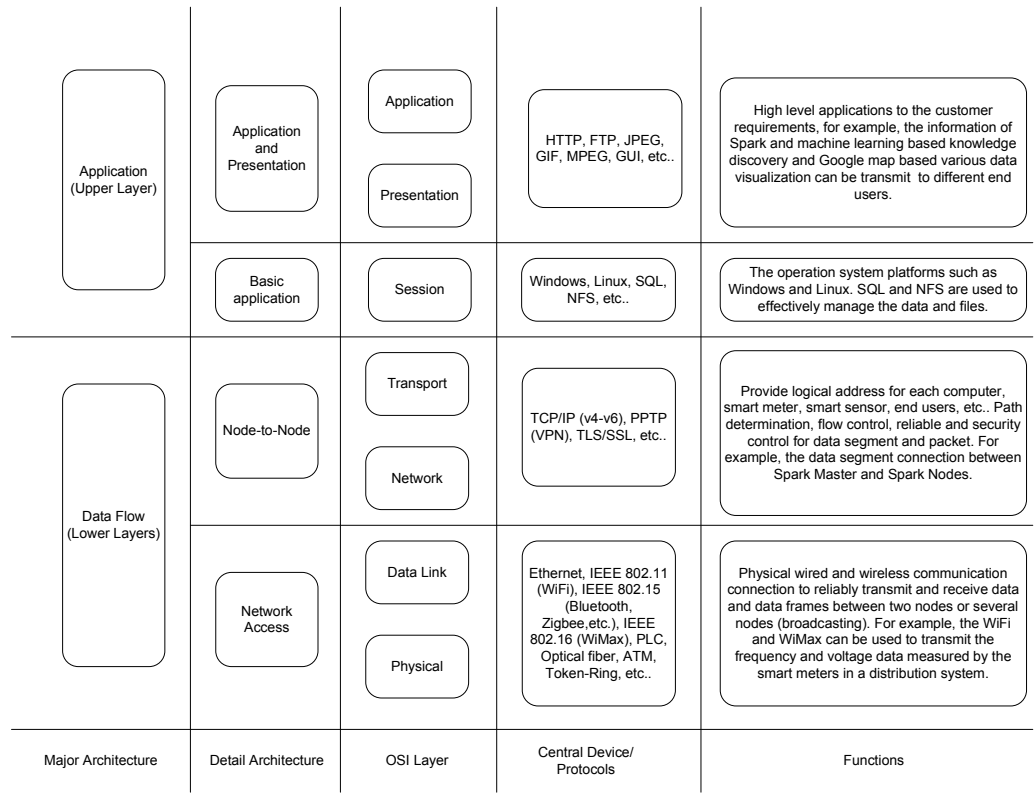

Figure 2.3: Architecture of the communication design.

2. **Node-to-Node**: The major task of this layer is to provide addresses for network access devices such as computers, smart sensors, servers, and end users, to send and receive data among the network access devices, determine the path of data flow, control the security and quality of service (QoS).

3. **Application and Presentation**: The major task of this layer is to implement the complex data structure on the data visualization platform.

## 2.2 The Spark Based Knowledge Discovery

In this paper, the Apche Spark based computer cluster is used to discover the hidden knowledge in the collected big data, which provides a resilient, fast, and effective parallel com-putation platform for many machine learning algorithms such as regression, classification, recognition, etc.. [75]. As shown in Fig 2.2, the Apache Spark contains: Spark core, Spark SQL, Spark Streaming, Spark GraphX, Spark Mlib.

### 2.2.1 Spark Core

The Spark Core contains a lot of basic functions such as task scheduling, memory management, fault recovery, etc. [35]. RDD is one of the most significant features of Spark core, which means many distributed computer nodes with databases can be organized and manipulated in parallel. RDD saves the application states such as check points periodically, which means the system and computation can be recovered quickly after some work nodes loss or failed. In this paper, considering the different requirements from different customers, multiple tasks can be computed in parallel with the high-effective platform. The big data processing results can be accessed in multiple end users, respectively.

### 2.2.2 Spark SQL

Spark SQL is an efficient package focusing on processing data, and supporting a lot of data forms [35]. In this paper, the Python is used as the programming language, which also supports a lot of data forms such as XLS, MAT, etc. [76].

### 2.2.3 Spark Streaming

Spark Streaming is a useful component for processing the live streaming of data [35]. Compared with Hadoop, which needs to frequently load and reload data from disk, Spark can save and store the data in memory, which dramatically increasing the data processing speed.

## 2.3 The Google Earth Based Data Visualization

Many open sources visualization tools are used to support a simple platform for the researchers to integrate any kind of data with their geospatial products [37, 77]. Both of the two platforms can be attached multiple layers with small editing. For Google Earth, the map and other information are displayed in KML format. Compared with arcGIS, Google Earth is more convenient and vivid for primer operations. Google Earth is considered as a better choice to build the power system model for DU campus in this paper.

## 2.4 Results

### 2.4.1 A Description of Demand Response in this method

The proposed platform can be used to satisfy a variety of different requirements for different users. Fig. 2.4 illustrates the demand response with the temperatures and the unit price in the SG of DU campus. In this model, the massive heterogeneous data such as temperature, electrical price, voltage, frequency, etc. are collected by the pervasively located smart sensors in real-time. Then the designed platform is used to compute the demand responses according to different time periods. The red and blue pyramids indicate the real-time temperature and electricity price, respectively. The green pyramid denotes the corresponding demand response in real-time.

### 2.4.2 A Description of Renewable Energy-PV

As shown in Fig. 2.5, this model illustrates the ratio between PV generation and total electrical power consumption. In the red circle, the three white bars denote the total power consumption of three areas in Ritchie Center building, the three red bars indicate the electrical power bought from utility, the white areas not covered by the red bars are the PV generations. As shown in Fig. 2.5, the PV generation is decreasing, and electrical power from utility is increasing in a cloudy day. With this real-time information, the customers, ISO, and power plant can schedule their behaviors and managements correspondingly.

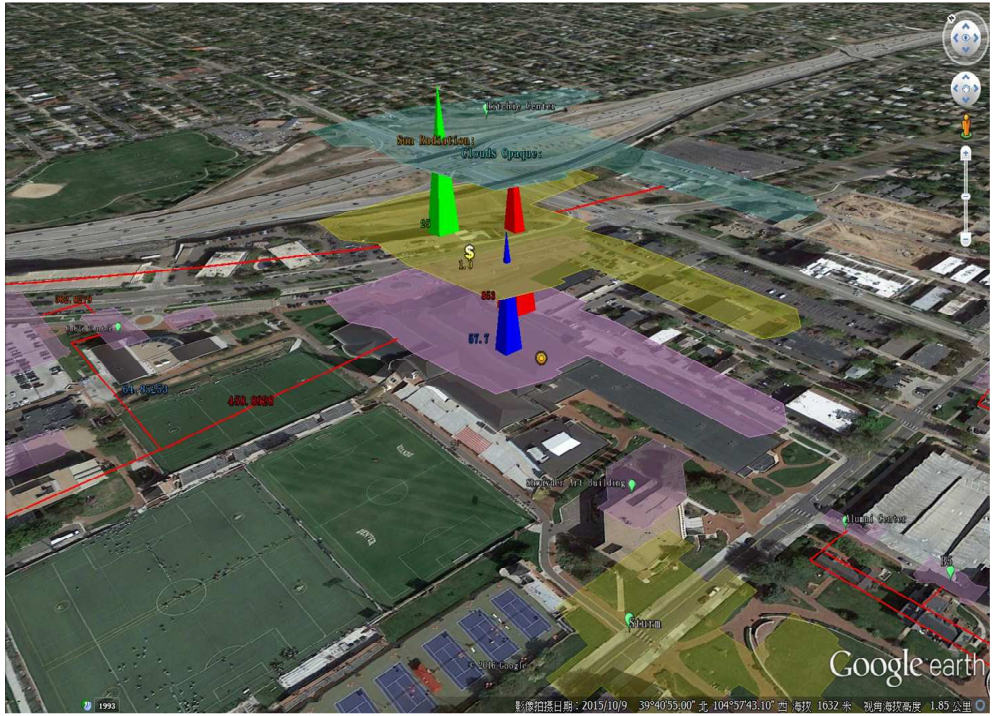

Figure 2.4: Real-time demand response in DU campus.

### 2.4.3 Multivariate Linear Regression Analysis

The multivariate linear regression is used to analyze the correlationships between response variable $Y_t$ and explanatory variables $X_t^i$, where $Y_t$ and $X_t^i$ are time series variables, and $i \in \{1, 2, 3, \cdots\}$ indicates the different types of explanatory variables, $t$ is the time index [78]. In this paper, the response variable $Y_t$ is the load profile of the Daniels College of Business building. The explanatory variables are collected from HVAC of the Daniels College of Business building including air pressure index 1 in HVAC ($X_t^1$), fan tuning index ($X_t^2$), wind tunnel temperature 1 ($X_t^3$), wind tunnel temperature 2 ($X_t^4$), and air pressure index 2 ($X_t^5$). The regression output is a time series variable $\hat{Y}_t$. The collected data are normalized

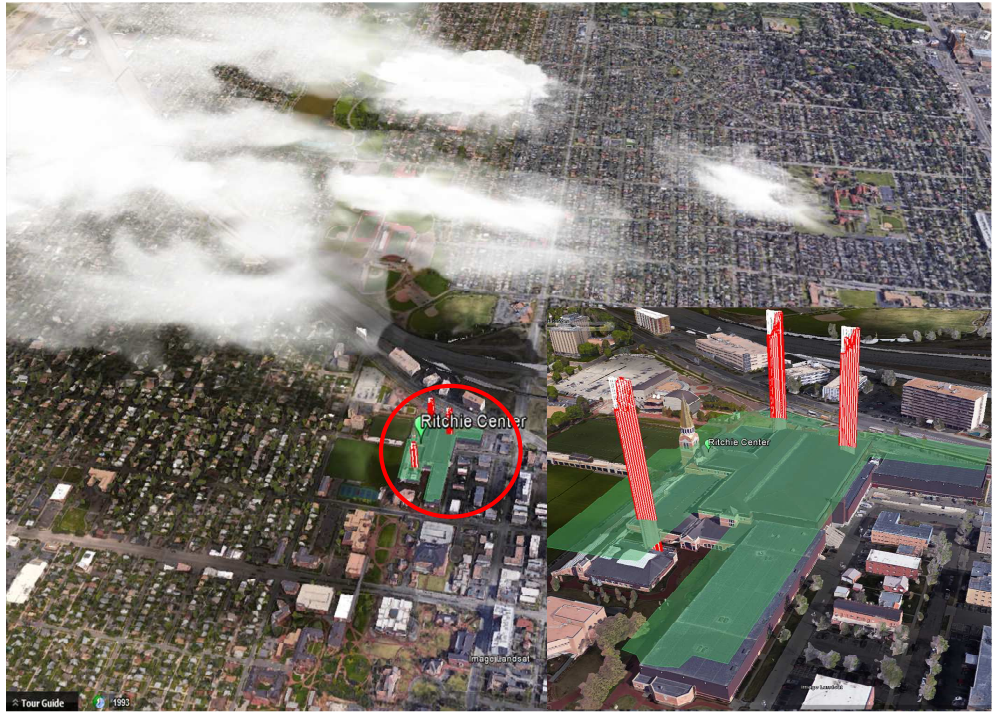

Figure 2.5: Real-time PV generation and total electrical power consumption in DU campus.

between 0 and 1.

Ordinary least squares (OLS) is a widely used to compute the coefficients of the explanatory variables, and optimize the squared differences between $Y_t$ and $\hat{Y}_t$. However, in the OLS, it is assumed that the error term is a constant variance, and ignores the heteroscedasticity, which concerns the expectations of the second moment of the errors [79]. This means the OLS is not a valid estimation approach. Considering the heteroscedasticity impact, the feasible generalized least squares (FGLS) can estimate the errors covariance matrix to improve the efficiency of the regression. Therefore, in this paper, the FGLS is used to compute the multivariate linear regression in big data analysis [79].

As shown in Table 2.1, compared to the original data without normalization, the regression error of the normalized data is much smaller. The regression errors of FGLS are less than the OLS in both original and normalized data regression. As shown in Table 2.2, the explanatory variable $X_t^4$, wind tunnel temperature 2, has the largest impact; and $X_t^1$, air pressure index 1, has the smallest impact to the response variable $Y_t$, the load profile of the Daniels College of Business building.

Table 2.1: Regression Performance (Squared Error)

|  | Original | Normalized |
|---|---|---|
| OLS | 635 | 0.457 |
| FGLS | 345 | 0.342 |

Table 2.2: The coefficients of explanatory variables with FGLS

|  | $X_t^1$ | $X_t^2$ | $X_t^3$ | $X_t^4$ | $X_t^5$ |
|---|---|---|---|---|---|
| Coefficient | 0.0251 | 0.0826 | 0.0857 | 0.1719 | 0.0318 |

# Chapter 3

# Load Forecasting with Support Vector Regression (SVR)

## 3.1 The Architecture of the Proposed Approach

In the left part of Fig. 3.1, the optimization of the hyper-parameters contains GTA and PSO to avoid be trapped into local minimum. The global solution space is spitted by the GTA into the local solution spaces. In the Map phase, because of the independency among the local solution spaces, they are traversed by GTA in parallel. In the Reduce phase, one or several local solution spaces are selected with the minimum training errors. In the second step, the selected local solution spaces are optimized by the PSO in the similar manner. The optimal parameters can be generated after comparison in the Reduce phase. Finally, the short-term load can be forecasted with the optimized hyper-parameters.

In the right part of Fig. 3.1, in the part of distribution system network reconfiguration, a three-phase balanced distribution system model is built with the forecasted load profiles. The SOCP is used to relax the three-phase balanced optimal power flow problem into a convex problem. In the Map phase, the ADMM is used to compute the three-phase balanced

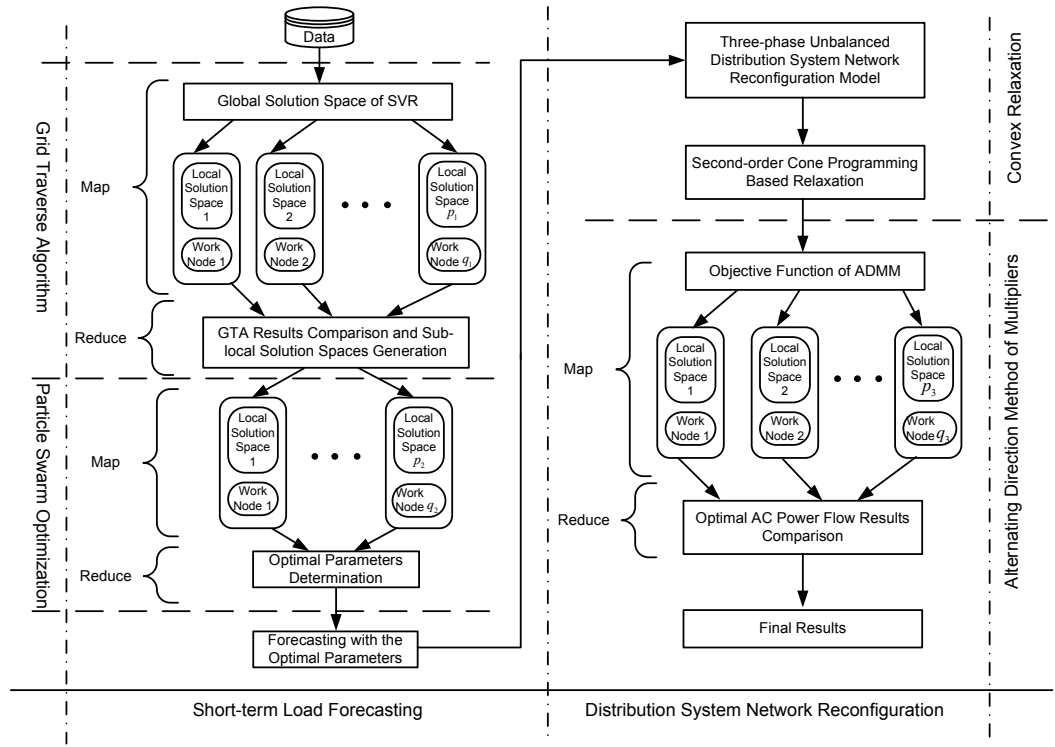

Figure 3.1: The flowchart of the proposed approach.

optimal power flow in parallel.

## 3.2 SVR-Based Short-term Load Forecaster

### 3.2.1 SVR Formulation

In this part, the SVR-based short-term load forecaster is trained by the collected historical data to get the optimal hyper-parameters. The objective function of the Kernel based SVR can be built to minimize the forecast error with the soft margin as follows:

$$R_{risk} = \min_{\varepsilon,\omega,\xi_{i_1},\xi_{i_1}^*,C,b,\gamma} \{\frac{1}{2}\omega^T\omega + C\sum_{i_1=1}^{n}(\xi_{i_1} + \xi_{i_1}^*)\} \qquad (3.2.1)$$

Subject to

$$\begin{cases} L'_{i_1} - f(x_{i_1}) \leqslant \varepsilon + \xi_{i_1}, \\ -L'_{i_1} + f(x_{i_1}) \leqslant \varepsilon + \xi_{i_1}^*, \\ \xi_{i_1}, \xi_{i_1}^* \geqslant 0. \end{cases} \qquad (3.2.2)$$

where in (3.2.1), $f$ is a Kernel based regression function, $i_1$ is a time index, $\frac{1}{2}\omega^T\omega$ indicates the flatness of the regression coefficients, the second item is the tube violation, $C$ is a trade-off coefficient between the first two items, $\xi_{i_1}$ and $\xi_{i_1}^*$ indicates the two training errors. In [47], the risk function (3.2.1) with the constraint (3.2.2) can be derived to a dual problem with Karush-Kuhn-Tucker (KKT) condition. However, the parameters $\gamma$, $C$, and $\varepsilon$ are still need to be determined, which are the critical factors to the performance of the forecaster [47]. The detail derivative of the SVM or SVR and its dual forms can be found in [47, 80]. Then, as shown in Fig. 3.1, a two-step based parameter optimization approach is designed to compute the optimal parameters.

### 3.2.2 Two-step parameter optimization

Because the parameters $\gamma$, $C$, and $\varepsilon$ cannot be solved with the convex optimization, they are be defined as hyper-parameters in [81]. There are several approaches are proposed for the hyper-parameters such as random search, sequential search, and Gaussian process [81, 82]. Considering the complexity and feasibility, a grid traverse search based two-step hyper-parameter optimization is proposed for the SVR based short-term load forecasting [48].

**First Step: the GTA Procedure**

As shown in Fig. 3.1, the GTA procedure is the first step for the hyper-parameters optimization, which aims to traverse the global solution space into one or several local solution spaces. In the second step, the local solution spaces can be searched with the PSO based approach. The proposed approach is based on the increasing computation capability and new computer cluster cooperation soughs, for example Mapreduce. In the first step, the three hyper-parameters are initialized with their upper bounds, lower bounds, and grid searching steps. Then, a traversing vector $\mathbf{H}$ can be generated as a finite multi-Cartesian product, which is critical for the Mapreduce process. $H_{j_2}$ is an element in $\mathbf{H}$. For each $H_{j_2}$, the loss function of SVR $R_{risk}$ can be computed independently. As shown in Fig. 3.1, they can be computed in parallel to reduce the computation time. In the last step, the minimum $R_{risk}$ is selected. In addition, if several $H_{j_2}$ are selected, all of them are transmitted to the second step for PSO optimization. In real-applications, if there is high requirement for the time consumption, the GTA can provide good enough parameter optimization.

---
**Algorithm 1** GTA for Hyper-parameter Optimization

    **Objective**: Shrink the global solution spaces into one or several local solution spaces.

    **Initialization**: Hyper-parameters initialization and multi-Cartesian product generation for the GTA.

    **Grid Traverse Searching**: For each core or each process, assign the $H_{j_2}$ to compute the $R_{risk}$, which can be computed in parallel with the Mapreduce model.

    **Determine Local Solution Space**: Collected all the results, and select the local solution spaces with minimum $R_{risk}$.

---

**Second Step: the PSO Procedure**

In this paper, the PSO procedure is designed as a "fine" optimization for the hyper-parameters, which can be implemented as the scenarios with less time consumption requirements.

After initialization of the particles, for particle $i_3$, its velocity and position can be com-

puted as following:

$$\boldsymbol{\nu}_{i_3}^{\Omega}(t) = \boldsymbol{\nu}_{i_3}^{\Omega}(t-1) + \varphi_1\theta_1(\boldsymbol{\eta}_{i_3}^{\Omega} - \boldsymbol{\alpha}_{i_3}^{\Omega}(t-1)) \quad (3.2.3a)$$
$$+ \varphi_2\theta_2(\boldsymbol{\eta}_g^{\Omega} - \boldsymbol{\alpha}_{i_3}^{\Omega}(t-1)),$$
$$\boldsymbol{\alpha}_{i_3}^{\Omega}(t) = \boldsymbol{\alpha}_{i_3}^{\Omega}(t-1) + \boldsymbol{\nu}_{i_3}^{\Omega}(t), \quad (3.2.3b)$$

where the acceleration coefficients are defined as $\varphi_1$ and $\varphi_2$, $\theta_1$ and $\theta_2$ can be seemed as two independently weightiness coefficients, the best historical position and the best position are defined as $\boldsymbol{\eta}_{i_3}^{\Omega}$ and $\boldsymbol{\eta}_g^{\Omega}$, respectively, and hyper-parameter $\Omega = [\gamma\ C\ \varepsilon]$. $\boldsymbol{\alpha}_{i_3}^{\Omega}(t)$ and $\boldsymbol{\nu}_{i_3}^{\Omega}(t)$ are the position and velocity vectors, respectively.

## 3.3 Distribution System Network Reconfiguration

The topology of a distribution system can be represented in a graph with buses and branches: $\mathcal{G} = [\mathcal{V}, \mathcal{E}]$. Then the branch flow model can be built as follows [56, 58]:

$$s_i = \sum_j S_{ij} - \sum_k (S_{ki} - l_{ki}z_{ki}), \quad (3.3.1a)$$
$$v_j = v_i - 2(r_{ij}P_{ij} + x_{ij}Q_{ij}) + (r_{ij}^2 + x_{ij}^2)l_{ij}, \quad (3.3.1b)$$
$$l_{ij} = (P_{ij}^2 + Q_{ij}^2)/v_j, \quad (3.3.1c)$$

where $l_{ij} := |I_{ij}|^2$, $v_i := |V_i|^2$, $S_{ij}$, $P_{ij}$, $Q_{ij}$ and $z_{ij}$ indicate the complex power flow, active power, reactive power, and impedance on branch $ij \in \mathcal{E}$, $S_{ij} = P_{ij} + \mathbf{i}Q_{ij}$, $P_{ij} = |I_{ij}|^2 r_{ij}$ and $z_{ij} = r_{ij} + \mathbf{i}x_{ij}$.

During the operation of network reconfigurations, the topology of the distribution system is keeping radial and avoid any loops. Considering the characteristics of the three-phase

Table 3.1: Results of Network Reconfiguration

| No. | Scenario | Bus No. | Opened Switches | Original $P_{Loss}$ | New $P_{Loss}$ | **Loss Reduction** |
|---|---|---|---|---|---|---|
| 1. | Load Increasing | 83 | TS-3, TS-4 | 54.2 kW | 36.7 kW | 32.28 % |
| 2. | Load Increasing | 300 | TS-2, TS-3 | 42.4 kW | 31.5 kW | 25.71 % |
| 3. | Load Increasing | 95 | TS-3, TS-4 | 78.5 kW | 67.0 kW | 14.65 % |
| 4. | Load Increasing | 49 | TS-1, TS-4 | 17.6 kW | 13.4 kW | 23.86 % |
| 5. | Load Decreasing | 47 | TS-2, TS-4 | 39.4 kW | 22.3 kW | 43.40 % |
| 6. | Load Decreasing | 108 | TS-1, TS-4 | 24.1 kW | 20.1 kW | 16.59 % |
| 7. | Load Decreasing | 250 | TS-3, TS-4 | 29.5 kW | 26.8 kW | 9.15 % |
| 8. | Load Decreasing | 56 | TS-1, TS-2 | 35.2 kW | 33.7 kW | 4.27 % |

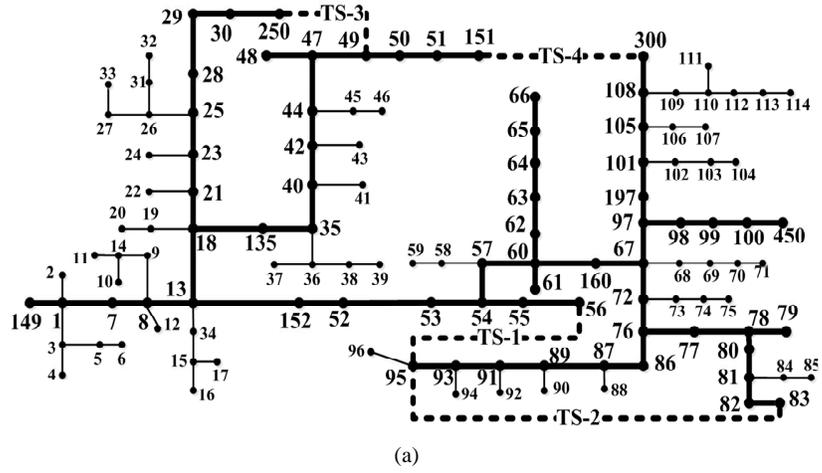

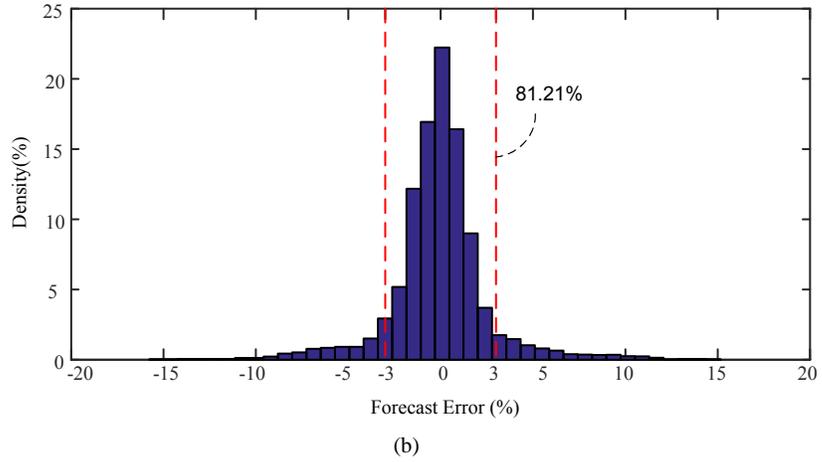

Figure 3.2: (a) The IEEE 123-bus based distribution system, (b) the error distribution of short-term load forecasting.

balanced distribution system, the SOCP relaxation inequalities can be represented as follows [56, 58]:

$$\frac{|S_{ij}|^2}{v_i} \leq l_{ij}, \qquad (3.3.2)$$

where (3.3.2) can be used to instead of (3.3.1c) as the inequalities constraints. In this paper,

the objective function is defined as total line loss as follows:

$$F = \sum_{\mathcal{E}} P_{ij}, \qquad (3.3.3)$$

where the constraints contain (3.3.1a), (3.3.1b), (3.3.2), and the basic physical constraints:

$$V_{i,min} \leq V_i \leq V_{i,max}, \qquad (3.3.4a)$$

$$I_{ij} \leq I_{ij,max}. \qquad (3.3.4b)$$

Considering the ADMM, the objective function (3.3.3) with the constraints (3.3.1a), (3.3.1b), (3.3.2), (3.3.4a), and (3.3.4b) can be decomposed into a dual problem. The detail derivatives of the ADMM can be found in [56, 58].

During the parallel traverse of all statuses of the switches, the topology of the distribution system is keeping as a radial network without any loops, which can be formulated as [38]:

$$\mathbf{rank}(A) = N - d, \qquad (3.3.5a)$$

$$\sum_{\mathcal{E}} a_{ij} = N - d, \qquad (3.3.5b)$$

$$\sum_{\mathcal{E}_k} a_{ij} = M_k - 1. \qquad (3.3.5c)$$

where $A$ is the adjacency matrix of the graph $\mathcal{G}$, $d$ is the number of slack bus, $N$ is the number of buses $\mathcal{V}$, $M_k$ is the number of branches in path $\mathcal{E}_k$, and $a_{ij}$ is an element of $A$:

$$a_{ij} = \begin{cases} 1, & if\ bus\ i\ and\ bus\ j\ are\ connected, \\ 0, & else. \end{cases} \qquad (3.3.6)$$

Considering the limited number of switches, the proposed approach is designed to traverse all the permutations and determine the optimal configuration of the distribution system.

For example, the modified IEEE 123-bus system with 4 switches indicates 16 scenarios with all the permutations of the switches [38]. With the topology constraints discussed above, the number of scenarios can be reduced in different scenarios. Then, considering the independency of each configuration (permutation), all the permutation can be implemented into different cores or processes and computed independently, which dramatically reduces the computation time and keeps the convexity to get the guaranteed optimization results.

## 3.4 Numerical Results

As shown in Fig. 3.2(a), the test bench is based on the IEEE 123-bus distribution system, and four initially opened tie switches TS-1, TS-2, TS-3 and TS-4 are added to make the system topology changeable, and the detail information can be found in [38, 83]. A test load data set is from a partner utility's distribution feeder.

### 3.4.1 Short-term Load Forecasting

The test load data contains four seasons of one year, and 30 days are selected for each season. The proposed SVR based load forecasting approach is used for 1-hour-ahead sliding window forecasting with 1 second resolution. The training data is 5 times as the test data. The distribution of the forecaster errors is shown in Fig. 3.2(b). The mean absolute precentage error (MAPE) is 2.23%, normalized root-mean-square error (NRMSE) is 4.03%, and more than 80% of the errors are accumulated between (-3.1%, 3.1%).

### 3.4.2 Network Reconfiguration

As shown in Table 3.1, considering the load increasing, for example, in scenario 1, with the forecasting results, there is a load increasing 20.31% in bus 83. The system loss reduces 32.28% with the proposed approach. For the load decreasing, for example, in scenario 5, with the forecasting results, there is a load decreasing 55.4% in bus 47. The system loss reduces 43.40% with the proposed approach. The average loss reduction for the load increasing scenarios is 24.13%, the average loss reduction for the load decreasing scenarios is 18.35%, and the total average for all scenarios is 21.24%.

Table 3.2: Performance comparison

| Methods | Loss Reduction | Computation Time (s) |
|---|---|---|
| GA based traditional | 17.87% | 107 |
| Proposed Approach | 21.24% | 30 |

### 3.4.3 Comparison

As shown in Table 3.2, compared with the traditional network reconfiguration approach with the genetic algorithm (GA), the proposed approach has less computation time and more loss reduction. Furthermore, the proposed approach is more intuitive, and convenient for implementation in different programming language such as python and Matlab.

# Chapter 4

# Multi-timescale power system optimization with renewable energy on two-levels

## 4.1 The Flowchart of the Proposed Approach

In this paper, multi-timescale stochastic algorithm [84] is applied to improve the operation cost of a distribution power system. We have proposed the configurable stochastic approach with the multi-timescale scheduling procedure, which consists of submodels including day-ahead power scheduling, real-time power trading constraint and OPF in the distribution system. The two parts stochastic optimization is implemented in this paper.

As in Fig. 4.1, the proposed approach consists of two parts, the stochastic optimization for hourly scheduling at the substation, and the three-phase unbalanced OPF for minutes operation at feeders. By the proposed two-part framework, the optimal day-ahead scheduling power purchased from the utilities for the next 24 hours is determined in the first part. In the second part, based on the results of the first part, the OPF is computed in minutes to reduce the system loss and the total system cost within an hour.

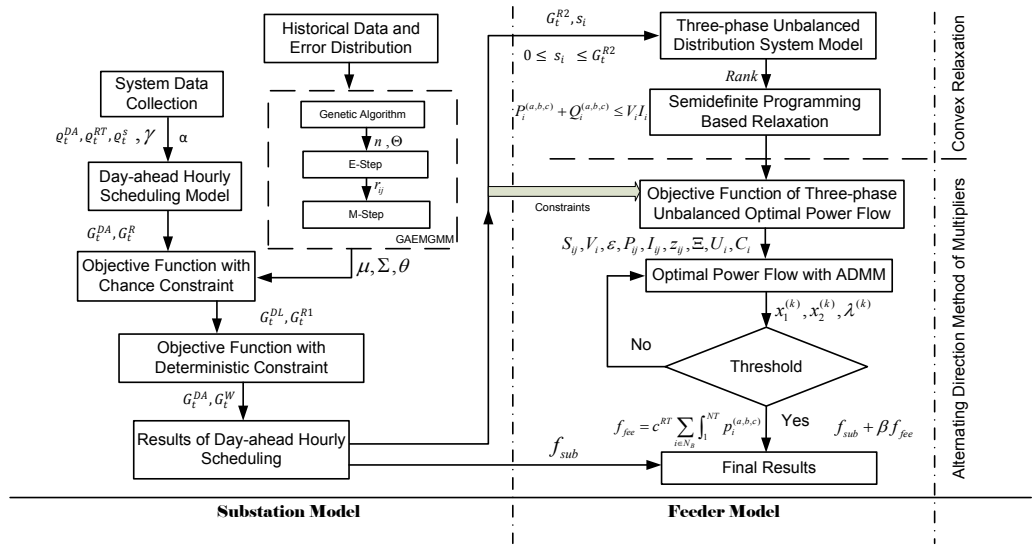

Figure 4.1: The flowchart of proposed approach.

In the left part of Fig. 4.1, the error distribution of the forecast result is given by the historical data of the distribution renewable generation. Combined with the day-ahead hourly scheduling, an objective function can be formulated with the chance constraints for the forecast errors. According to GAEMGMM, the error distribution model can be accurately modeled with several Gaussian components. And the chance constraints can be formulated into the deterministic forms for the stochastic optimization. Finally, the optimal hourly schedule can be determined and the optimal operation cost is computed at the substation level.

In the right part of Fig. 4.1, at feeder level, a three-phase unbalanced optimal power flow is used to model and compute the distribution system loss in minutes level. In this paper, the renewable generation such as micro wind turbines and PV panels derated, $2.5\%$ power are reserved for the OPF regulation, which aims to minimize the system loss. Considering the nonconvexity of the three-phase unbalanced OPF, SDP is used to relax the problem. Then, the objective function of system loss can be solved with ADMM to further reduce computation time. Finally, the three-phase unbalanced OPF can be computed to minimize the system loss at feeder level.

The total cost is formulated in (4.1.1), which can be presented as the sum of the operation cost at the substation level and feeder level:

$$C_{total} = \min(f_{sub} + \beta f_{fee}) \tag{4.1.1}$$

where the cost of substation and feeder levels are ($f_{sub}$) and ($f_{fee}$), respectively, and $\beta$ is a weightness coefficient. Given three different ways to working with $\beta$, the operation cost at the substation level (feeder level) is attacking more attention if $\beta > 1$ ($\beta < 1$). $\beta$ is defined as 1, when the two levels are the same important.

## 4.2 Day-ahead hourly scheduling at the substation level

### 4.2.1 Problem formulation

The load consumption cost model at the substation level can be defined as:

$$f_{sub} = \sum_{t=1}^{T} \left( (\varrho_t^{DA} G_t^{DA} + \varrho_t^{R} G_t^{R}) + (\delta_t \cdot \varrho_t^{RT} G_t^{RT}) \right. \\ \left. + (1 - \delta_t) \cdot \varrho_t^{s} G_t^{RT} \right) \tag{4.2.1}$$

where $t = \{1, 2, \cdots, T\}$ represents the time intervals, $t$ indicates the $t$th time interval and $T$ is used to describe how many time intervals we have in this situation. $\varrho_t^{DA} G_t^{DA}$ is the generation cost of day-ahead scheduling and $\varrho_t^{R} G_t^{R}$ is the renewable generation cost. $\delta_t \cdot \varrho_t^{RT} G_t^{RT}$ presents the deviation power purchased from RT market to satisfy the power balance equation. And $(1 - \delta_t) \cdot \varrho_t^{s} G_t^{RT}$ is the corrective action (CA), which indicates the redundant energy will be reselled to the market in a lower price. It is noticed that $G^{R1}$ is used to supply the power consumption at the substation level for the first part, which is defined as $97.5\%$ of the total renewable power generation for each hour in this paper.

A typical forecasting error model represented by a normal distribution is used to incorporate the uncertainty of renewable generation. Here, the load forecasting error model is also represented as normal distribution. The study of the data forecasting is not the major concentrate in this paper, and the hourly forecasted renewable generation and the hourly load forecast are described as:

$$G^R = G_f^R + G_f^R * G_{err1} \tag{4.2.2}$$

$$G^{DL} = G_f^{DL} + G_f^{DL} * G_{err2} \tag{4.2.3}$$

subject to:

$$G_t^{DA} + G_t^{RT} + G_t^W = G_t^{DL} \tag{4.2.4a}$$

$$\varrho_t^s < \varrho_t^R < \varrho_t^{DA} < \varrho_t^{RT} \tag{4.2.4b}$$

$$G^{RT,min} \leq G_t^{RT} \leq G^{RT,max} \tag{4.2.4c}$$

$$G^{DA,min} \leq G_t^{DA} \leq G^{DA,max} \tag{4.2.4d}$$

$$G^{R1,min} \leq G_t^{R1} \leq G^{R1,max} \tag{4.2.4e}$$

$$\delta_t = \begin{cases} 1 & G_t^{DA} + G_t^W \leqslant G_t^{DL} \\ 0 & G_t^{DA} + G_t^W > G_t^{DL} \end{cases} \tag{4.2.4f}$$

$$G_t^{RT} \begin{cases} \geqslant 0 & buy\ power\ from\ bulk\ system \\ < 0 & sell\ power\ to\ bulk\ system \end{cases} \quad (4.2.4g)$$

$$Pr(G_t^{DL} \leqslant G_t^{DA} + G_t^W) \geqslant \gamma \quad (4.2.4h)$$

$$Pr(\rho G_t^{R1} - G_t^W \leqslant 0) \geqslant \alpha \quad (4.2.4i)$$

where the constraints include the modified power balanced equation, market price limitation, capacity limits for $G_t^{RT}, G_t^{DA}, G_t^{R1}$. $\delta_t$ is a binary variable to describe the relationship between $G_t^{DA} + G_t^W$ and $G_t^{DL}$. We assume the relationship of the unit prices in (4.2.4b), the RT price ($\varrho_t^{RT}$) is the highest, and $\varrho_t^R$ is cheaper than $\varrho_t^{DA}$. It is easy to understand that the $\varrho_t^s$ should be the lowest in case of purchasing the redundant power generation from the day-ahead market. (4.2.4h) defines that the load demand will not exceed the sum amount of the renewable generation and the day-ahead power scheduling with a prescribed probability $\gamma$. In (4.2.4i), for each hour, the amount of the used renewable generation for all buses should be larger than the renewable generation at chance $\alpha$, where $0 < \rho < 100\%$.

### 4.2.2 Genetic-based Expectation Maximization Algorithm for learning Gaussian Mixture Model

A brief description of the finite GAEMGMM is given below. Then, it is used to model the forecasting error distribution of the renewable generation.

GMM is an unique form of the finite mixture model. For the finite mixture model in (4.2.5), it is the sum of more than one components ($N > 1$) with different weights ($\epsilon_n$) in

$\mathcal{R}^q$, which indicates:

$$p(x|\Theta) = \sum_{n=1}^{N} \epsilon_n p(x|\theta_n) \tag{4.2.5}$$

The weights in (4.2.5) can be calculated in [85] and has the intuitive interpretation to be non-negative ($\epsilon_n \geqslant 0$), and the sum equals to 1. Each component in the model obeys to a normal distribution, which is restricted by $\theta_n = (\mu_n, \mathbf{\Sigma}_n)$, the matrix of the means vector and the covariance.

The EM algorithm is used as the standard approach to calculate the parameters of the mixture model, which consists of an expectation-step (E-step) and a maximization-step (M-step).

For E-step, the complete data $\mathcal{Z} = \{(x_1, \xi_1), (x_2, \xi_2), \cdots, (x_m, \xi_m)\}$, where $\{x_1, x_2, \cdots, x_m\}$ is known as the observed data and incomplete, $\xi_m$ is the component identity of $x_m$. The algorithm will be ended when the log likelihood function in (4.2.6) reaches the convergence.

$$\mathcal{L}(x|\Theta) = \sum_{m=1}^{M} \log \left( \sum_{n=1}^{N} \xi_n(x_m|\theta_n) \right) \tag{4.2.6}$$

The posterior probability ($\xi_n$) at the $l$-th iteration is computed as:

$$\xi_n^{(l)} = \frac{\epsilon_n^{(l)} p(x_m|\theta_m)}{\sum_{n=1}^{N} \epsilon_n^{(l)} p(x_m|\theta_n^{(l)})} \tag{4.2.7}$$

For M-step, the parameters of the GMM is reestimated by $\xi_n^{(l)}$.

$$\epsilon_n^{(l+1)} = \frac{\sum_{m=1}^{M} \xi_{m,n}}{m} \tag{4.2.8}$$

$$\mu_n^{l+1} = \frac{\sum_{m=1}^{M} \xi_{m,n} X_m}{\sum_{m=1}^{M} \xi_{m,n}} \tag{4.2.9}$$

$$\Sigma_n^{(l+1)} = \frac{\sum_{m=1}^{M} \xi_{m,n}(x_m - \mu_n^{(l+1)})(x_m - \mu_n^{(l+1)})}{\sum_{m=1}^{M} \xi_{m,n}} \tag{4.2.10}$$

For learning the GMM in [68], the GAEM shows its superiority on selecting the number of components based on the minimun description length (MDL) criterion. In [86], the MDL criterion is widely used on selection. Compared with the standard EM, the new algorithm is less sensitive to the initialization. GAEM is capable to explore the parameter more thoroughly, because of its population-based search skill. Meanwhile, the new algorithm still remain the property of monotonic convergence as before.

A comparison with the GSM, GMM, and the GAEMGMM is illustrated as shown in Fig. 4.2. Fig. 4.2(b) displays the modeled error distribution result by GSM, which cannot modeled the aggregated error distribution accurately. The results of standard GMM with 3 components are shown in Fig. 4.2(c) and Fig. 4.2(d). Because the standard GMM cannot determine the number of the components automatically, which also bring some errors in the error distribution modeling. The GAEMGMM are employed to model the original error distribution in Fig. 4.2(e) and Fig. 4.2(f), which can automatically determine the number of the components and reduce the modeling error in this process.

As in (4.2.11), the ratio of the residual deviation $\eta$ is used to evaluate the performances of different approaches

$$\eta = \frac{(\nabla_d - \nabla_{org})^2}{\nabla_{org}^2} \cdot 100\% \tag{4.2.11}$$

where $\nabla_{org}$ is the envelope of the original forecasting error distribution, and $\nabla_d$ is the envelope of the deformed forecasting error model by GSM, general GMM and GAEMGMM. According to (4.2.11), the ratio of the residual deviation of GSM is 2.17%, GMM is 1.27%, the GAEMGMM is 0.13%. It is clear that the proposed approach has the best performance for error distribution modeling.

### 4.2.3 Chance Constraint

According to discussion above, the proposed GAEMGMM is capable to accurately model the aggregated error distribution and determine the components number at each feeder, which can effectively convert the chance-constraint of the forecasting error model in (4.2.4h) (4.2.4i) into a deterministic problem.

Because $G_{err1}$ and $G_{err2}$ is normally distributed with mean $(\mu_1, \mu_2)$, covariance $(\Sigma_1)$ for multiple renewable generators and variance $(\sigma_2^2)$ for load forecast. The joint probability distribution of the renewable generators can be obtained by the proposed GAEMGMM model with a high accuracy, which demonstrates the efficiency of the proposed approach further more.

Based on the chance-constraint in (4.2.4h) and (4.2.4i), it is assumed that

$$y_1 = G^{DL} - G^{DA} - G^W \tag{4.2.12}$$

$$y_2 = \rho G^{R1} - G^W \tag{4.2.13}$$

The expectation and the variance can be calculated as:

$$E(y_1) = G_f(1+\mu_2)^{DL} - G^{DA} - G^W \tag{4.2.14}$$

$$V(y_1) = G_f^{DL} * \sigma_2 \tag{4.2.15}$$

$$E(y_2) = \rho G_f^{R1}(1+\mu_1) - G^W \tag{4.2.16}$$

$$V(y_2) = G_f^{R^T} \Sigma_1 G_f^R \tag{4.2.17}$$

According to this, the chance-constraint in (4.2.4h), (4.2.4i) can be converted in a deterministic formulation as (4.2.18) and (4.2.19) with the proposed GMM model.

$$Pr(y_1 \leqslant 0) = \Phi\Big(\frac{0 - E(y_1)}{V(y_1)}\Big) \geqslant \gamma \qquad (4.2.18)$$

$$Pr(y_2 \leqslant 0) = \Phi\Big(\frac{0 - E(y_2)}{V(y_2)}\Big) \geqslant \alpha \qquad (4.2.19)$$

Where $\Phi(\cdot)$ indicates the cumulative distribution function of the standard normal distribution. Taking the inverse of $\Phi(\cdot)$, rewriting the resulting the following analytical formulation:

$$G^W + G^{DA} \geqslant \Phi^{-1}(\gamma)\sigma_2 G_f^{DL} + G_f^{DL} + \mu_2 G_f^{DL} \qquad (4.2.20)$$

$$G^W \geqslant \Phi^{-1}(\alpha)\rho\big((G_f^R)^T \Sigma_1 G_f^R\big)^{\frac{1}{2}} + \rho G_f^{R1} + \rho\mu_1 G_f^{R1} \qquad (4.2.21)$$

## 4.3 Optimal power flow (OPF) and the ADMM based Semidefinite Programming (SDP) relaxation

### 4.3.1 ADMM

Along with blending the decomposability of dual decomposition, ADMM shows the superior convergence properties of augmented Lagrangians [72]. For a general ADMM problem, the optimization problem is formulated as:

$$\min_x f_1(x_1) + f_2(x_2) \qquad (4.3.1)$$

subject to:

$$\begin{aligned} A_1 x_1 + A_2 x_2 &= b \\ x_1 \in \mathcal{K}_1, x_2 &\in \mathcal{K}_2 \end{aligned} \qquad (4.3.2)$$

$\mathcal{K}_1, \mathcal{K}_2$ are defined as convex sets. Then the objective is augmented as (the constraint is as (4.3.2)):

$$\min_x f_1(x_1) + f_2(x_2) + \frac{\rho}{2} \parallel A_1 x_1 + A_2 x_2 - b \parallel_2^2 \tag{4.3.3}$$

$\rho$ is a constant and never less than zero, which is used to decide if the objective is a augmented ($\rho > 0$) or standard ($\rho = 0$) Lagrangian. Then $\lambda$ is generally defined as the Lagrange multiplier for the equality constraint in (4.3.2).

After the formula deformation, ADMM repeats the iterations, for $k = 1, 2, 3 \cdots$

$$x_1^{(k)} = \arg\min_{x_1} L_\rho(x_1, x_2^{(k-1)}, \lambda^{(k-1)}), \tag{4.3.4a}$$

$$x_2^{(k)} = \arg\min_{x_2} L_\rho(x_1^{(k)}, x_2, \lambda^{k-1}), \tag{4.3.4b}$$

$$\lambda^{(k)} = \lambda^{(k-1)} + \rho(A_1 x_1^{(k)} + A_2 x_2^{(k)} - b). \tag{4.3.4c}$$

Refer to [72], ADMM is guaranteed to reach the converged optimal solution with less restricted conditions.

### 4.3.2 ADMM in this paper

This paper aims to minimize the total operation cost of the distribution power system, in other words, the proposed three-phase unbalanced network can reduce the operation cost.

A distributed network is modeled as a tree radial topology $\mathcal{G}$, which indicates that each bus $i$ in this system only has one parent bus $U_i$ and a few children buses $C_i$. $\mathcal{G} = (\mathcal{N}_B, \mathcal{N}_l)$, where $\mathcal{N}_B = \{0, 1, \cdots, n\}$ represents the set of buses. $\mathcal{N}_l$ is the set of the distributed lines, which are used to connect the buses in $\mathcal{N}_B$. Each $i \in \mathcal{N}_L = \{1, 2, \cdots, n\}$ indicates the line connected from bus $i$ to its parent bus $U_i$.

Compared with bus injection model, branch power flow provides much more use for distribution network and stable computation results. For a three-phase power system, the branch power flow $\mathbf{S}_i^{(a,b,c)}$, complex branch current $\mathbf{I}_i^{(a,b,c)}$, voltage magnitude $\mathbf{V}_i^{(a,b,c)}$ and complex impedance $\mathbf{\Omega}_i^{(a,b,c)}$ are defined as follows, which expressed as instantaneous space vectors and indicates the power flow from bus $i$ to its parent bus $U_i$:

$$\mathbf{S}_i = \begin{bmatrix} S_i^a \\ S_i^b \\ S_i^c \end{bmatrix} = \begin{bmatrix} P_i^a \\ P_i^b \\ P_i^c \end{bmatrix} + \mathbf{i} \begin{bmatrix} Q_i^a \\ Q_i^b \\ Q_i^c \end{bmatrix} \quad (4.3.5)$$

$$\mathbf{I}_i = \begin{bmatrix} |I_i^a|^2 \\ |I_i^b|^2 \\ |I_i^c|^2 \end{bmatrix} \quad (4.3.6)$$

$$\mathbf{V}_i = \begin{bmatrix} |V_i^a|^2 \\ |V_i^b|^2 \\ |V_i^c|^2 \end{bmatrix} \quad (4.3.7)$$

$$\mathbf{\Omega}_i = \begin{bmatrix} \Omega_i^a \\ \Omega_i^b \\ \Omega_i^c \end{bmatrix} = \begin{bmatrix} r_i^a \\ r_i^b \\ r_i^c \end{bmatrix} + \mathbf{i} \begin{bmatrix} x_i^a \\ x_i^b \\ x_i^c \end{bmatrix} \quad (4.3.8)$$

According to the definition of branch flow model, the network $\mathcal{G}$ is defined as:

$$\begin{aligned} V_{U_i}^{(a,b,c)} =& \mathbf{V}_i - 2(r_i^{(a,b,c)} P_i^{(a,b,c)} + x_i^{(a,b,c)} Q_i^{(a,b,c)}) \\ & + \mathbf{I}_i((r_i^{(a,b,c)})^2 + (x_i^{(a,b,c)})^2) \end{aligned} \quad (4.3.9)$$

$$P_i^{(a,b,c)} + Q_i^{(a,b,c)} = \mathbf{V}_i \mathbf{I}_i \quad (4.3.10)$$

$$\sum_{j \in C_i} (P_j^{(a,b,c)}) - \mathbf{I}_i r_i^{(a,b,c)} + p_i^{(a,b,c)} = P_i^{(a,b,c)} \quad (4.3.11)$$

$$\sum_{j \in C_i} (Q_j^{(a,b,c)}) - \mathbf{I}_i x_i^{(a,b,c)} + q_i^{(a,b,c)} = Q_i^{(a,b,c)} \quad (4.3.12)$$

$$\begin{pmatrix} \mathbf{V}_i & \mathbf{S}_i \\ \mathbf{S}_i^H & \mathbf{I}_i \end{pmatrix} \in \mathbb{S}_+ \qquad (4.3.13)$$

$$rank \begin{pmatrix} \mathbf{V}_i & \mathbf{S}_i \\ \mathbf{S}_i^H & \mathbf{I}_i \end{pmatrix} = 1 \qquad (4.3.14)$$

Where $P$ and $Q$ ($p$ and $q$) indicates the bus $i \in \mathcal{N}_L$ ($i \in \mathcal{N}_B$). $S_0$ is the root of the topology with no parent bus. $\mathbb{S}$ and $\mathbb{S}_+$ are used to describe the hermitian and the positive semidefinite matrix, respectively. And $(.)^H$ denotes the hermitian transpose of the matrix. Because the proposed distributed system is radial with unique phase angle of the current and voltage at each bus, the branch flow model in (4.3.9)-(4.3.12) can be regarded as a complete AC power flow now [72].

The objective function is formulated as:

$$f_{fee} = \varrho^{RT} \sum_{i \in \mathcal{N}_B} \int_1^{NT} p_i^{(a,b,c)} \qquad (4.3.15)$$

subject to: (4.3.9), (4.3.10), (4.3.11), (4.3.12), (4.3.13), (4.3.14), (4.3.16a), (4.3.16b) and (4.3.17).

The OPF problem in (4.3.15) cannot be regarded as a convex problem because of the rank constraint in (4.3.14). Due to the SDP relaxations in [87], the rank constraint in (4.3.14) can be removed and obtain a lower bound for the revised OPF problem in (4.3.18). It has been improved in [88], the semidefinite relaxed OPF is exact if the optimal solution of (4.3.18) still satisfy the rank constraint and the original OPF problem is also optimal.

$$\mathbf{V}_{\mathbf{i}min} \leqslant \mathbf{V}_{\mathbf{i}} \leqslant \mathbf{V}_{\mathbf{i}max}, i \in \mathcal{N}_B \qquad (4.3.16a)$$

$$s_i^{(a,b,c)} = p_i^{(a,b,c)} + \mathbf{i}q_i^{(a,b,c)}. \qquad (4.3.16b)$$

$$0 \leqslant s_i^{(a,b,c)} \leqslant G^{R2} \tag{4.3.17}$$

In (4.3.16a), the magnitude of the voltage obeys to a reasonable range. The system controlling parameter $s_i^{(a,b,c)}$ is used to benefit on reducing the system loss, which is defined to be provided by the renewable generation and limited in (4.3.17). Based on the schedule results of first step, $G^{R2}$ is defined as $2.5\%$ of the total renewable generation and used at feeder level for step two.

The relaxation OPF formulation is summarized as:

$$f_{fee} = \varrho^{RT} \sum_{i \in \mathcal{N}_B} \int_1^{NT} p_i^{(a,b,c)} \tag{4.3.18}$$

subject to: (4.3.9), (4.3.10), (4.3.11), (4.3.12), (4.3.13), (4.3.16a), (4.3.16b) and (4.3.17).

## 4.4 Numerical Simulation and Results

### 4.4.1 Error Distribution Modeling Comparison

The predicted data, actual data, and error distribution are provided by National Renewable Energy Laboratory (NREL) [89], which can be modeled with different approaches as following. In table. 4.1, the proposed method can obtain a better ratio of the residual deviation $\eta$, as defined in (4.2.11), than others.

### 4.4.2 Day-Ahead Dispatching Cost at the Substation Model

The numerical results for evaluating the proposed method are tested based on the IEEE 123-bus system. Four wind turbines (100kWh for each) are connected at bus 25, 35, 76 and 105, respectively. The PV panels (400kWh in total) are installed at bus 28, 47, 49, 57, 64, 93 and 97, which are used to demonstrate that the hybrid power system can work efficiently and reliably with the proposed stochastic approach.

Table 4.1: Comparison on $\eta$ for different fitting model

| $\eta$ (%) | Spring | Summer | Autumn | Winter |
|---|---|---|---|---|
| GAEMGMM | 0.09 | 0.13 | 0.16 | 0.15 |
| General GMM | 1.03 | 1.27 | 1.44 | 1.23 |
| GSM | 2.39 | 2.17 | 2.15 | 2.10 |

The hourly renewable generation with the total operation cost is simulated in Fig. 4.3(a), where the blue (yellow) bars indicates the wind (solar) power generation hourly in a day. In this case, the lower limit of the forecasting error model in chance-constraint (4.2.4h) is set to 97.0%.

Fig. 4.3(a) describes the total operation cost with and without CA, which indicate with red and green curve, respectively. As shown in Fig. 4.3(a), in Golden, Colorado, a typical day with a windy night and a sunny daytime is selected with 24 hours data. The peak generation of the wind turbines (blue bars) and PVs (yellow bars) occur at midnight and 14:00, respectively. The red line represents the total operating cost with CA, which indicates the system can resell the redundant power at a lower price $\varrho_t^s$ to the current electricity market. Then, the total operation cost can be reduced with the benefit from the resell. The total operation costs at 13:00 and 14:00 for both scenarios (with and without CA) are the same, which indicates there are no redundant energy to resell and the CA doesnt occur. The similar scenarios also occur at 7:00 and 18:00. In the rest time, the redundant energy is resold to the market with a lower price to reduce the total operation cost.

$$a = \frac{G_t^{RT}}{G_t^{DL}} * 100\%, \tag{4.4.1a}$$

$$a = \frac{G_t^{DA} + G_t^{R}}{G_t^{DL}} * 100\%, \tag{4.4.1b}$$

The proposed approach is compared with the GSM based approach in Fig. 4.3(b) for 24 hours. For each hour, the left bar describes the GSM fitted model and its corresponding cost is displayed as the green dashed line. The right bar describes the GA-EM fitted model and its corresponding cost is displayed as the red dashed line. It is clear that the GA-EM fitted model contains higher accuracy, the corresponding system operation cost is lower than the system operation cost of the GSM fitted model.

Specifically, in the left bar (with GSM fitted model), the yellow bar denotes the $G_t^{RT}$ (in percent, can be computed as (4.4.1) as following) and the rest is $G_t^{DA} + G_t^{R}$. Similarly, in the right bar, the orange and dark blue bars also indicate the $G_t^{RT}$ and $G_t^{DA} + G_t^{R}$ with the GA-EM fitted model. It is clearly that the orange bar is shorter than the yellow bar, which indicates the GA-EM fitted model contains less errors than the GSM fitted model, and the system requires less energy from the RT market.

**Case.1 Different Operation Costs for One Year Data at Substation Level.**

Based on the discussion above, one year data with four seasons are employed to validate the proposed approach. Specifically, for each season, 30 days data are selected, and we also employed the GMM and GSM to compare. As shown in Table 4.2, the proposed approach contains the minimum cost for spring, summer, autumn, and winter. It is demonstrated that with the high accuracy fitted model, the corresponding operation cost can be reduced significantly, which also indicates the related fuel and carbon emission can be saved.

**Case.2 Total Operation Cost with Different Percentage Limit in Chance Constraint.**

The operating cost with the $\gamma$ of $95.0\%$ is lower than others. The highest cost is obtained with $\gamma = 99.0\%$. This illustrates that the higher reliability can result in higher total operation cost. This is due to that a bigger $\gamma$ places a stricter constraint that $G_t^{DA} + G_t^{W}$ is higher or equal to the $G_t^{DL}$ in (4.2.4h). On the contrary, the operating cost decrease if we define a smaller $\gamma$ for the system.

As shown in 4.4, the chance constraint probability $\gamma$ in (4.2.4h) is set as $95.0\%$, $97.0\%$, $99.0\%$ with blue, red, and orange color, respectively. It is clearly that the operating cost with $\gamma = 95.0\%$ is the lowest among others, and the highest cost is obtained with $\gamma = 99.0\%$. These results illustrate that a high chance constraint probability $\gamma$ requires a high demand of $G_t^{DA}$ in equation (4.2.4h), which indicates the operation cost is also increasing.

Table 4.2: Comparisons based on Fig. 4.2 at substation level

| Season | Test Days | Average Cost for Each Day($) | | |
|---|---|---|---|---|
| | | Proposed Method | GMM(General) | GSM |
| Spring | 30 | 3345.24 | 3662.58 | 4102.56 |
| Summer | 30 | 5170.86 | 5556.21 | 6321.99 |
| Autumn | 30 | 3123.97 | 3400.02 | 4112.66 |
| Winter | 30 | 5456.02 | 5822.21 | 6363.42 |

Table 4.3: Time Consumption Comparison

| Method | IEEE 13-bus | IEEE 34-bus | IEEE 123-bus |
|---|---|---|---|
| Genetic Algorithm | 210.77 s | 243.39 s | 507.53 s |
| Interior-Point | 7.72 s | 11.33 s | 27.67 s |
| Proposed Method | 1.59 s | 3.72 s | 10.27 s |

On the contrary, with a low chance constraint probability $\gamma$, the corresponding operation cost is low.

### 4.4.3 Optimal Power Flow with ADMM

The genetic algorithm (GA) is an artificial intelligence algorithm to simulate natural evolutionary processes, through retaining a population of candidate solutions to search for the optimal one. Some techniques are used to create candidate, which is inspired by crossover and selection. Genetic algorithm is usually implemented as a computer simulation method, which denotes that, for an optimization problem, the abstract representation of a number of candidate solutions use this algorithm to evolve toward better solutions. The evolution starts with completely random individual populations. In each generation, the fitness of the all creatures from the current generation is evaluated, multiple individual(creatures) is randomly selected (based on their fitness) and generate new population through natural se-lection, the new generated population will repeat the same procedure to generate the next generation until satisfying the requirement of the system [90–92]. Considering the GA approach implemented in [93, 94], the population size is chosen as 600, which is enough to generate the new generation population with pinpoint accuracy. The probabilities of performing crossover and mutation are 0.8 and 0.08, respectively. The Matlab Global Optimization Toolbox and Distributed Evolutionary Algorithms in Python are used for the GA approach.

Table 4.4: Line Loss Comparison with or without OPF

| kWh | Spring | Summer | Autumn | Winter |
|---|---|---|---|---|
| Line Loss with OPF | 227 | 324 | 226 | 521 |
| Line Loss without OPF | 316 | 432 | 307 | 657 |

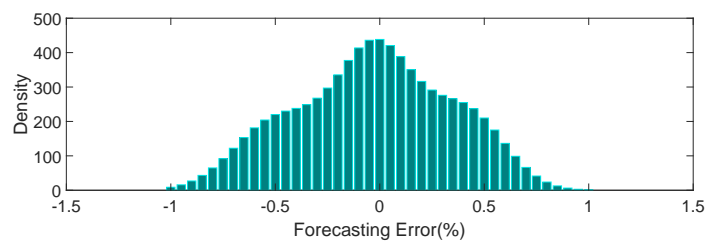

(a)

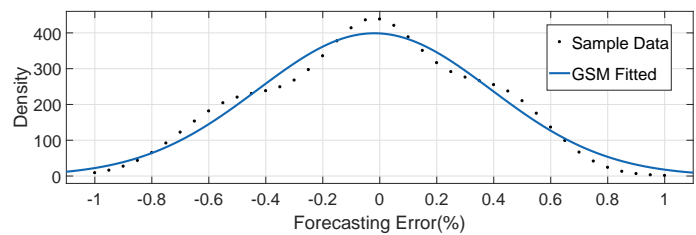

(b)

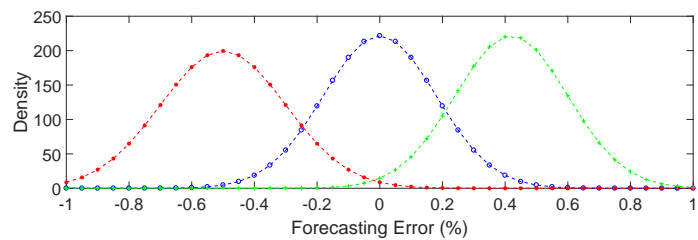

(c)

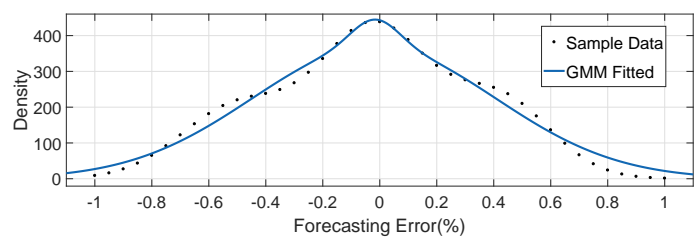

(d)

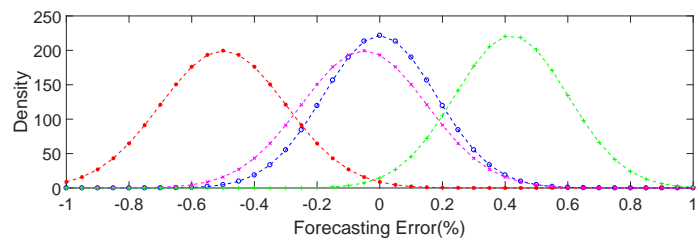

(e)

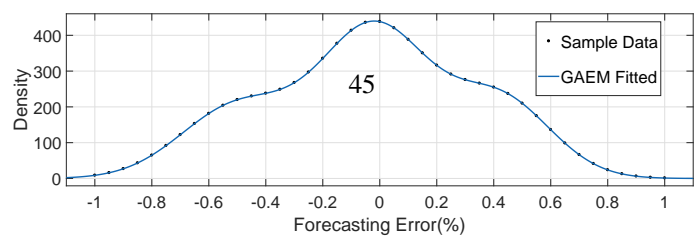

(f)

Figure 4.2: Comparisons between GSM, GMM and GAEMGMM: (a) the original aggregated error distribution, (b) the modeled result with GSM and comparing with the original original aggregated error distribution, (c) the original aggregated error distribution is mod-

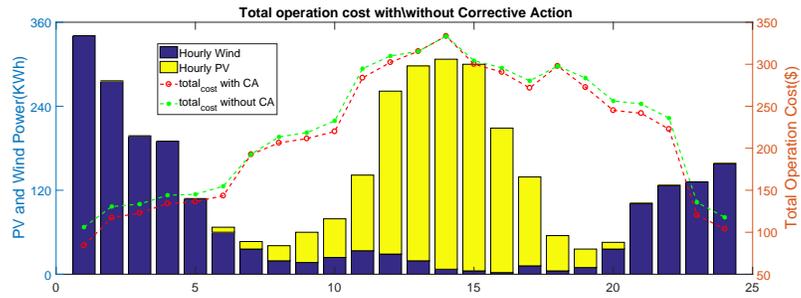

(a)

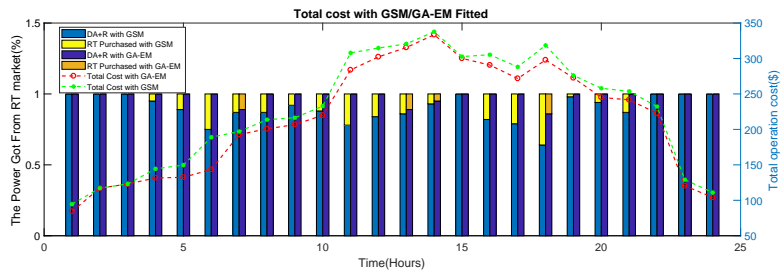

(b)

Figure 4.3: (a) Total Operation Cost with and without the CA. (b) Total Cost with GSM based approach and proposed approach.

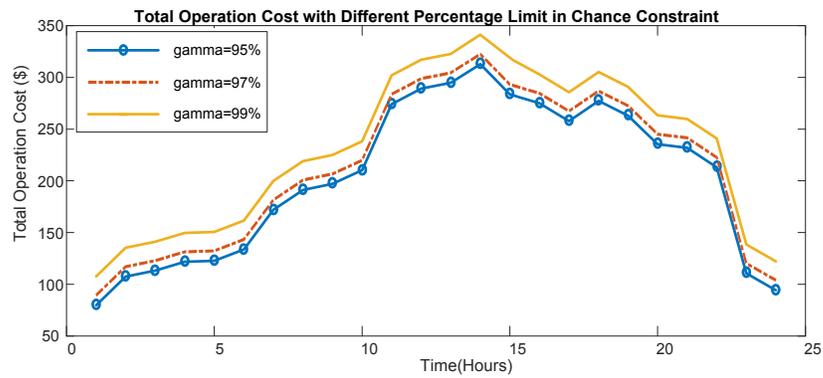

Figure 4.4: Total Operation Cost with Different $\gamma$ in Chance Constraint.

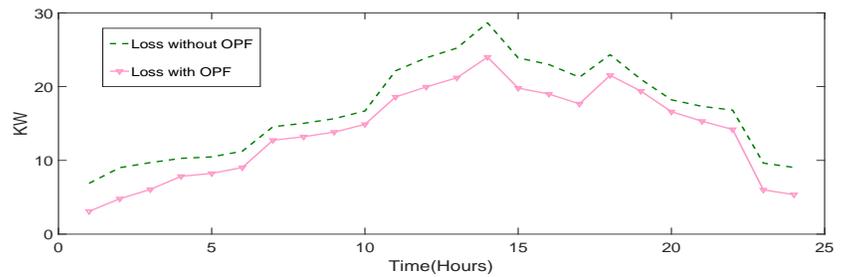

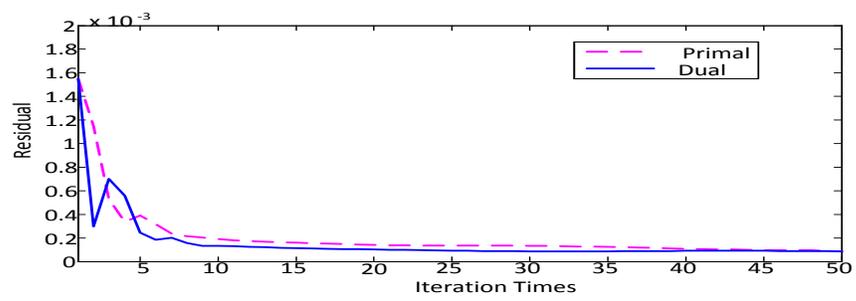

Figure 4.5: (a) The Line Loss of the Distribution System. (b) Convergency Analysis.

## Chapter 5

# Conclusion

A big data visualization platform is designed and built in this paper, which aims to discover the hidden useful knowledge for SG operation, control, and situation awareness. With Apache Spark, a computer cluster is built to discover the hidden useful knowledge in parallel. The Google Earth based visualization platform is used to visualize the discovered knowledge in various ways. In addition, the customers, ISOs, and power plants can send different requirements to the platform and implement different responses according to the received knowledge from the platform.

A short-term load forecasting based network reconfiguration is proposed to reduce the distribution system loss dynamically. Instead of the static load measurements at the scheduled time spots, the short-term load forecasting approach can provide the accurate future load profiles, which contains more information for the network reconfiguration. With the SOCP, the OPF of the three-phase balanced distribution system can be relaxed into a convex problem and solved with ADMM. The optimal reconfiguration is generated by the parallel computation with traversing all the permutations of the switche status. The whole proposed approach is designed as a distributed computation approach.

For the distribution system operator, a multi-timescale approach is proposed for a three-phase unbalanced distribution system. Considering the day-ahead dispatch with the substa-tion level, the objective is to minimize the system cost with renewable generation, and the chance constraint is build based on GMM and GAEM. For the feeder level, the objective also aims to minimize the system loss with the feeder level, an OPF problem is formulated for the three-phase unbalanced system, relaxed with SDP, and solved with ADMM. In the proposed multi-timescales model, the feeder scheduling is designed with higher time resolution and update frequencies, which can obtain near RT calculation. The numerical results demonstrate the effectiveness and validity of the proposed approach on real system (DU campus), medium-scale power system (IEEE 123-bus distribution system) and large-scale power system (IEEE 8500-bus distribution system) with different seasons, the operation cost of each of them can be improved with the proposed approach.

In the next step, with the increasing penetration of renewable energies, the wind and solar power will be considered, and other system such as traffic system with electrical vehicles will be also considered in the future research. Power reserves of the redundant power generated will be considered as an important role to further reduce the operation cost for the future research.